\documentclass[letter]{aa}

\usepackage{graphicx}
\usepackage{txfonts}

\newcommand{\Msun}{M$_{\odot}$}

\newcommand{\Mstar}{M$_{\star}$\ }

\begin{document}

   \title{Discovery of a jet from the single HAe/Be star HD 100546}

   \author{P. C. Schneider. \inst{1}
    \and  
           C. Dougados \inst{2}
    \and
          E. T. Whelan\inst{3}           
    \and
          J. Eisl\"offel \inst{4}
    \and 
           H. M. G\"unther \inst{5}
    \and
           N. Hu\'elamo \inst{6}
    \and
          I. Mendigut\'ia \inst{6}
    \and 
           R. D. Oudmaijer\inst{7}
    \and 
          Tracy L. Beck\inst{8}       
}

   \institute{Hamburger Sternwarte, Gojenbergsweg 112, D-21029 Hamburg, Germany \email{astro@pcschneider.eu}
     \and  
   Univ. Grenoble Alpes, CNRS, IPAG, 38000 Grenoble, France
        \and
    Maynooth University Department of Experimental Physics, National University of Ireland Maynooth, Maynooth Co. Kildare, Ireland
     \and
   Th\"uringer Landessternwarte, Sternwarte 5, D-07778 Tautenburg, Germany
      \and
   Massachusetts Institute of Technology, Kavli Institute for Astrophysics \& Space Research, 77 Massachusetts Avenue, Cambridge, MA 02139, USA
   \and 
   Centro de Astrobiología (CSIC-INTA), Departamento de Astrof\'isica, ESA-ESAC Campus, PO Box 78,  28692, Villanueva de la Ca\~{n}ada, Madrid, Spain
        \and
        School of Physics and Astronomy, University of Leeds, Leeds LS2 9JT, UK
    \and
      Space Telescope Science Institute, 3700 San Martin Drive, Baltimore, MD 21218, USA
}

\date{}
\abstract{Young accreting stars drive outflows that collimate into jets, which can be seen hundreds of au from their driving sources. Accretion and outflow activity cease with system age, and it is believed that magneto-centrifugally launched disk winds are critical agents in regulating accretion through the protoplanetary disk. Protostellar jets are well studied in classical T~Tauri stars ($M_\star\lesssim2\,M_\odot$), while few nearby ($d\lesssim150\,$pc) intermediate-mass stars ($M_\star=2-10\,M_\odot$), known as Herbig Ae/Be stars, have detected jets. We report VLT/MUSE observations of the Herbig~Ae/Be star HD~100546 and the discovery of a protostellar jet. The jet is similar in appearance to jets driven by low-mass stars and compares well with the jet of HD~163296, the only other known optical jet from a nearby Herbig~Ae/Be star. We derive a (one-sided) mass-loss rate in the jet of $\log \dot{M}_{jet} \sim -9.5$ (in $M_\odot$\,yr$^{-1}$) and a ratio of
outflow to accretion of roughly $3\times10^{-3}$, which is lower than that of CTTS jets.

The discovery of the HD~100546 jet is particularly interesting because the protoplanetary disk around HD~100546 shows a large radial gap, spiral structure, and might host a protoplanetary system. A bar-like
structure previously seen in H$\alpha$ with VLT/SPHERE shares the jet position angle, likely represents the base of the jet, and suggests a jet-launching region within about 2\,au. We conclude that the evolution of the disk at radii beyond a few au does not affect the ability of the system to launch jets. 
}
\maketitle
%

\section{Introduction}
Jets are integral to the star formation process, and much of what is understood about astrophysical jets in general comes from studies of young stellar objects (YSOs). Protostellar jets are observed across a wide mass range and in all pre-main sequence evolutionary stages \citep{Frank_2014}. Most studies of protostellar jets focused on classical T Tauri stars (CTTSs, class II low-mass YSOs) so that relatively little is known about jets at lower \citep[][]{Whelan2014} and higher stellar masses, that is, for 
the intermediate-mass Herbig Ae/Be stars ($M_\star\sim2-10\,M_\odot$). \citet{Corcoran_1998} reported that the flux ratio between [O~{\sc i}] and H$\alpha$ emission in HAe/Be stars follows the same trend as in CTTSs, where a tight connection between accretion and outflow rate has been observed \citep{Cabrit_1992, Hartigan_1995, Gunther2012}. Magnetic fields play a crucial role here, and magneto-centrifugal  launching  is accepted  as  the  likely mechanism  for  the  generation  of  protostellar jets \citep{Frank_2014}, which are launched from the inner disk. Jet-launching is also thought 
to affect the planet-forming accretion disk  \citep{Baruteau_2014, Dougados2018}.

The focus of this letter is the single Herbig Ae/Be star HD~100546 \citep[d = 109 $\pm$ 4~pc, \Mstar = 1.9 $\pm$ 0.1~\Msun;][respectively]{Lindegren_2018, Fairlamb_2015}. Herbig Ae/Be stars typically lack strong stellar magnetic fields \citep{Alecian_2013, Jaervinen_2019}, like HD~100546 \citep[$B_Z=89\pm26\,$G compared to kG fields in CTTSs;][]{Hubrig_2009} so that a transition from magnetically funneled accretion to perhaps boundary layer accretion could be expected for these stars and would likely also affect the jet-driving mechanism \citep{Ellerbroek2014}. 
HD~100546 is one of the nearest Herbig~Ae/Be stars and is particularly interesting because there is at least one planetary candidate in the system \citep[e.g.,][]{Brittain_2014, Currie_2015, Quanz_2015, Rameau_2017, Cugno2019}. Its protoplanetary disk shows spiral structure \citep{Boccaletti_2013,Follette_2017} and a large inner cavity, which makes HD~100546 a so-called pre-transitional disk system \citep{Brittain_2009,Mulders_2013, Walsh_2017}. HD~100546 had not been reported as driving a jet previous to this work, but asymmetric structures have been reported in the inner disk, such as a bar-like structure in H$\alpha$
\citep[5-10\,au][]{Mendigutia_2017} and a potential disk wind in SO \citep{Booth_2018} and Si~{\sc iii} \citep{Grady_2004}.
We present optical integral field spectroscopic observations of HD~100546, which reveal a jet that is launched perpendicular to the disk major axis in the typical jet lines.
Integral field spectroscopy is particularly well suited to the detection and analysis of protostellar jets. Specifically, it allows for the efficient isolation of jet features against contaminating effects such as the Point Spread Function (PSF)   wings or background emission and
kinemato-morphological studies, and  its field of view matches typical sizes of protostellar jets well.

\section{Observations, data reduction, and analysis}

Observations were performed with the European Southern Observatory (ESO) Multi-Unit Spectroscopic Explorer \citep[MUSE;][]{Bacon_2010}  located on the ESO Very Large Telescope (VLT). 
 Data for HD~100546 were 
obtained on 2017 May 4 (seeing$\sim 0\farcs65$, airmass $\sim1.4$).
The exposure times were consecutively increased during the observing run to balance observing efficiency and saturation effects that were due to the brightness of the central star. Images and spectra presented here are based on data 
obtained with $t_{exp}=90\,$s. The few exposures with
(much) shorter integration times provide no significant improvement.

\begin{figure}
  \centering
  \includegraphics[width=0.49\textwidth]{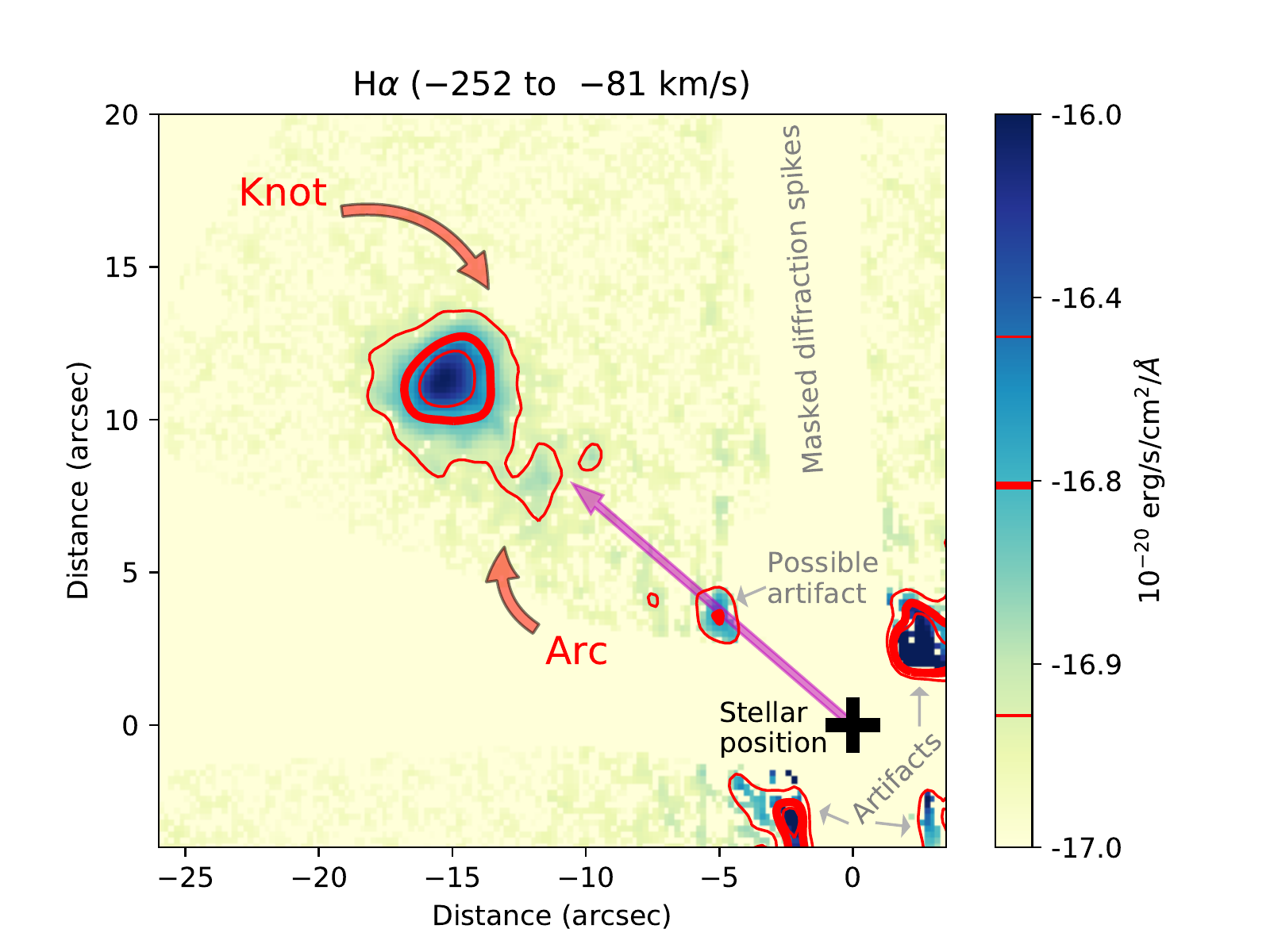}
  \caption{Majority of the field of view around HD~100546 in H$\alpha$. Stellar emission has been removed, and subtraction residuals are indicated as artifacts.}
  \label{fig:overview}
\end{figure}

Esoreflex \citep{Freudling_2013} and the MUSE pipeline version 2.2 were used for the data reduction. Because the jet emission is fainter than that of the central star, a PSF subtraction was performed using dedicated python routines, as described in Appendix~\ref{sect:datareduction}. Still, the main jet features are already visible before PSF subtraction (see Fig.~\ref{fig:full_overview}). All quoted velocities in this work refer to the stellar rest frame.

\section{Results \label{sect:results}}

\begin{figure}[t]
  \centering
  \includegraphics[width=0.49\textwidth]{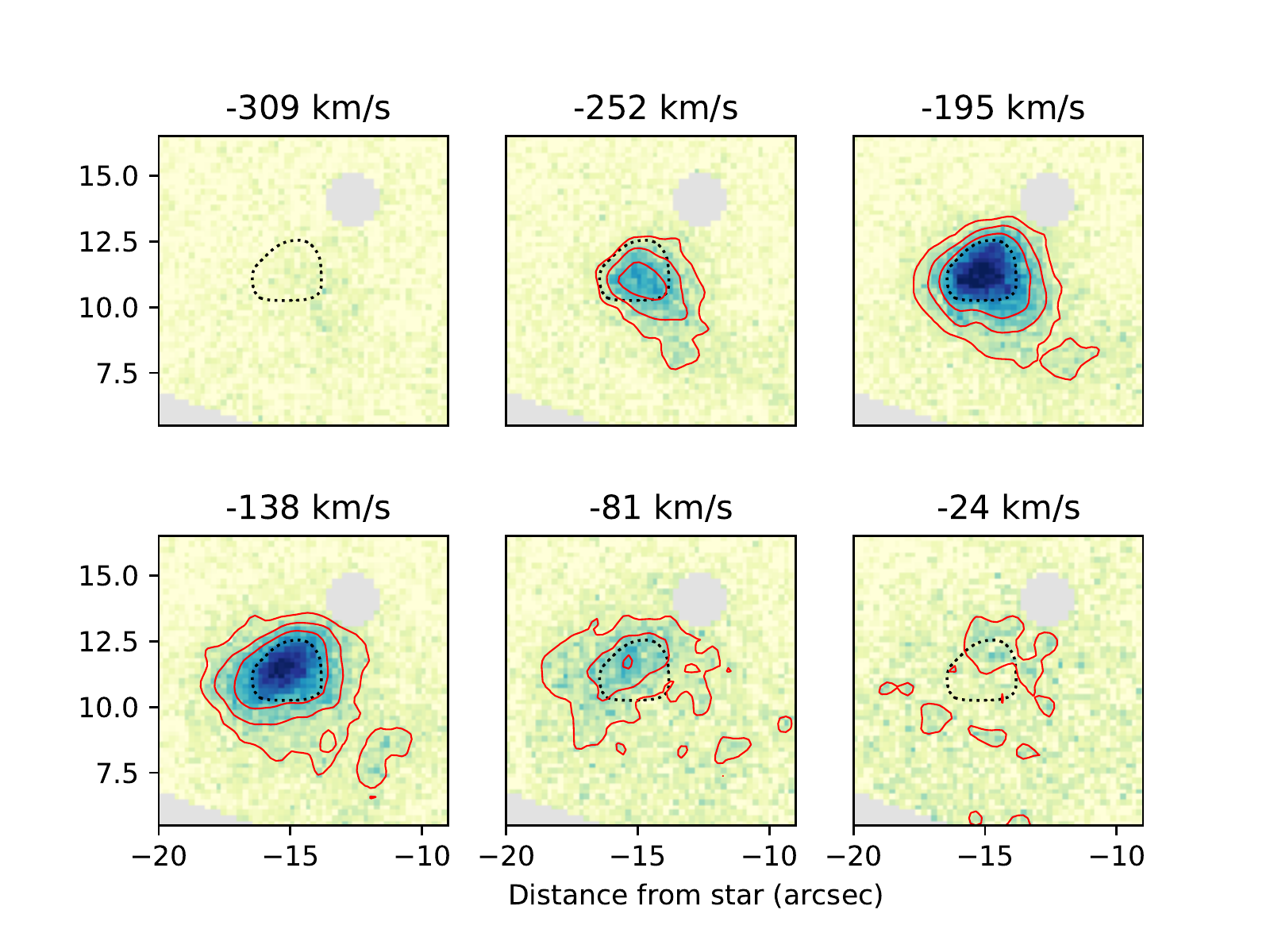}
  \caption{H$\alpha$ images of the jet in different velocity channels. To provide a reference between the different velocity channels, the dotted black contour displayed in all panels describes the flux integrated between -252 and -81 km\,s$^{-1}$.}
  \label{fig:zoom}
\end{figure}

Diffuse emission in typical emission lines associated with protostellar jets is clearly detected in the vicinity of HD~100546; see Fig.~\ref{fig:overview}. This emission
consists of an emission knot at a projected distance of 18.4\,arcsec (2005\,au) from the star (upper left quarter in Fig.~\ref{fig:overview}), and its radius is $r\approx2.5$\,arcsec where the flux dropped to one-third of the peak value. In addition, there is an arc-like structure closer to the star (below the knot).
The line that connects the knot and the arc-like structure points toward HD~100546 and implies a position angle of $53^\circ$ (E of N). The channel maps (Fig.~\ref{fig:zoom}) show more details of the emission. Specifically, they show that the mean (projected) plasma velocity at the knot is about -170\,km\,s$^{-1}$, which agrees  with  that of the arc. In  the -138\,km\,s$^{-1}$ channel, this structure has a banana-like shape: the wings bend backward toward the driving source, which is reminiscent of a typical bow shock. Assuming that the lateral extent of the arc-like structure represents an upper limit on the width of the jet, we find an opening angle of about $10^\circ$ for the jet; a similar value is found for the knot.

\begin{figure*}[t]
  \begin{centering}
  \includegraphics[width=0.89\textwidth]{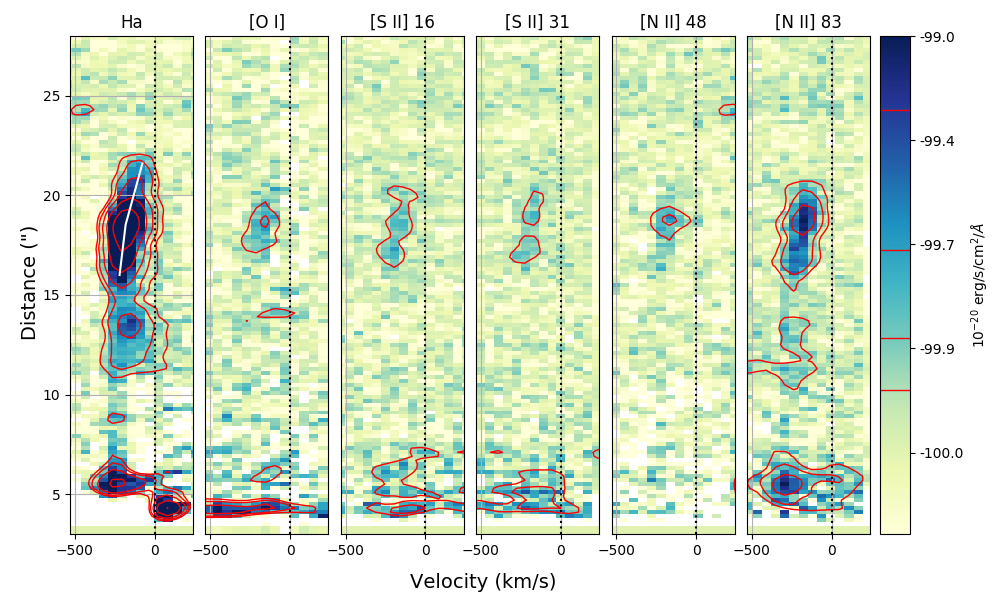}
  \caption{Position-velocity diagrams of the main jet emission lines. From left to right: H$\alpha$, [O~{\sc i}] $\lambda6300$, [S~{\sc ii}] $\lambda6716$, [S~{\sc ii}] $\lambda6731$, [N~{\sc ii}] $\lambda6548$, and [N~{\sc ii}] $\lambda6583$. The white line in the H$\alpha$ panel indicates the peak velocity.
  \label{fig:PVDs}}
  \end{centering}
\end{figure*}

The pseudo-longslit spectrum extracted along PA=53$^{\circ}$ reveals emission in the typical jet lines  consistently at a distance of 18\,arcsec from HD~100546 and at a velocity of about -170\,km\,s$^{-1}$ (Fig.~\ref{fig:PVDs}).
In addition, line emission is detected all the way down to 6~arcsec from the star, where the signal is lost in the PSF wings of the central star (especially in H$\alpha$ and [N~{\sc ii}]~$\lambda$6583).

The position-velocity diagrams (PVDs)  also show the arc-like structure (at 14\,arcsec). For the knot at 18.4\,arcsec, the velocity of the emission continuously decreases with increasing distance from the star (seen in all lines shown in Fig.~\ref{fig:PVDs}). In H$\alpha,$ the velocity is $-190$\,km\,s$^{-1}$ at a distance of 16" and just $-110$\,km\,s$^{-1}$ at 21". A similar trend is seen in [N~{\sc ii}]~$\lambda$6583 and is likely present in the other lines, although the signal-to-noise ratio (S/N) is too low to allow firm conclusions for them. With the observed distances and velocities, the dynamical age of the jet is 60 to 90\,years, assuming that the jet is launched perpendicular to the disk  \citep[with this assumption, the inclination with respect to the sky is either 46$^\circ$ or 57$^\circ$, see ][for a discussion of the disk inclination]{Mendigutia_2015,Mendigutia_2017}.

\begin{figure}[t]
 \sidecaption
  \includegraphics[width=0.49\textwidth]{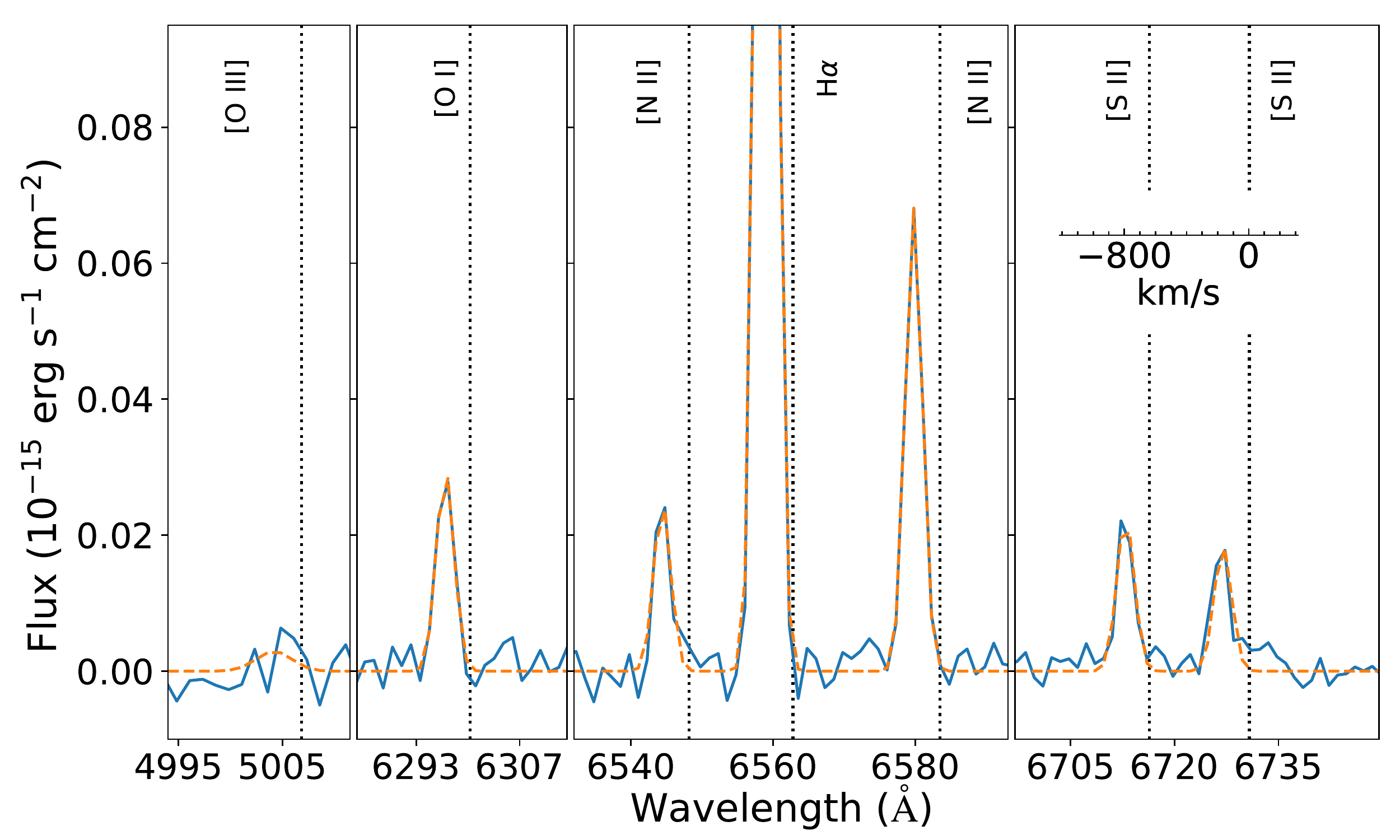}
  \caption{Selected spectral regions extracted at the knot. The axis scaling is similar in all panels in velocity and flux, i.e., the velocity scale shown in the [S~{\sc ii}] panel applies to all four panels equally.}
  \label{fig:spec}
\end{figure}

The line profiles extracted from the region centered on the H$\alpha$ peak, specifically from the region enclosed by the thick contour in Fig.~\ref{fig:overview},  are shown in Fig.~\ref{fig:spec}, and the measured line properties are provided in Table~\ref{tab:lines} \citep[the central star is not subject to a significant amount of extinction so that we did not deredden the measured fluxes; see][]{Fairlamb_2015}.
The best-fit line shifts are between -160 and -173~km\,s$^{-1}$. 
The wavelength regions around other potential jet lines show no significant emission; an example is the [O~{\sc iii}]~$\lambda5007$ line, 
which is expected to be strong in highly excited or ionized plasma, that is, for high shock velocities. Other lines also have similar upper limits as [O~{\sc iii}]. The line ratio between the two [S~{\sc ii}] lines around 6720\,\AA{} is density sensitive, and we derive a nominal electron density of $n_e\approx250\,$cm$^{-3}$ \citep[the 1\,$\sigma$ confidence range is $n_e<1200$\,cm$^{-3}$ from inspection of Fig.~1 in ][and the lower value is compatible with the low-density limit of the {[S~{\sc ii}]} ratio so that we cannot formally exclude low densities]{Bacciotti_1995}.

The line ratios, in particular [N~{\sc ii}]~$\lambda6583$ to [S~{\sc ii}]~$\lambda\lambda6716, 6731$ and [N {\sc ii}]~$\lambda6583$ to [O~{\sc i}]~$\lambda6300$  (see Table~\ref{tab:lines}), suggest shock velocities of $\sim80$\,km\,s$^{-1}$ rather than low-velocity (<50\,km\,s$^{-1}$) shocks, which would for instance have  [N {\sc ii}]~$\lambda6583$ to [O~{\sc i}]~$\lambda6300$ of $\lesssim1.0$ for low jet densities \citep[$n\leq 1,000\,$cm$^{-3}$, see Fig. 11 in ][]{Hartigan_1994}. Interestingly, the ratios of H$\alpha$ to H$\beta$ and of [S~{\sc ii}] $\lambda6731$ to $\lambda6716$  appear more compatible with the \citet{Hartigan_1994} models that have a preshock magnetic field of $\gtrsim100\,\mu$G.

To provide an estimate of the mass fluxes through the jet, we followed 
\citet{Hartigan_1994}. Specifically,  we used their Eq. 10, that is,
\begin{eqnarray}
\dot M  & = & 5.95\times10^{-8} \left(\frac{10^3\textnormal{
cm}^{-3}}{n_e}\right)\left(\frac{L_{6300}}{10^{-4} L_{\sun}}\right)
\nonumber\\
 & & \times \left(\frac{v_{sky}}{100\textnormal{ km
s}^{-1}}\right)\left(\frac{l_{sky}}{10^{16}\textnormal{ cm}}\right)^{-1}
M_{\sun} \textnormal{ yr}^{-1} \,,
\end{eqnarray}
where $L_{6300}$, $v_{sky}$, and $l_{sky}$ are the [\ion{O}{i}] line flux, projected jet velocity, and aperture length, respectively.

Our fluxes pertain to an aperture with $\approx1.5$\,arcsec radius, so that the projected length on the sky $l_{sky}$ is 327\,au ($4.9\times10^{15}$\,cm). With a mean projected 
velocity $v$ of 170\,km\,s$^{-1}$, we find a mass flow of 
\begin{equation}
  \dot{M}_{jet} =  2.8\times10^{-10} \left(\frac{250\textnormal{ cm}^{-3}}{n_e}\right) M_{\sun} \textnormal{ yr}^{-1} 
\end{equation}
through the blueshifted lobe for $L_{\textnormal{[O I]}} = 3.35\times10^{-8}\,L_\odot$. Outer streamlines that do not contribute to the shock emission or lower densities would increase the mass loss, while
other reasonable numerical constants or methods described in Appendix~\ref{sect:alt_mass_loss} provide similar values. We therefore conclude that a mass-loss rate around $\log \dot{M}_{jet}\sim-9.5$ is very reasonable for the HD~100546 jet.
Compared to an  accretion rate of $\dot{M}_{acc}\approx10^{-7}\,M_\odot\,$yr$^{-1}$ \citep[][]{Mendigutia_2015, Fairlamb_2015}, the above values give an outflow-to-accretion rate that is a few times $10^{-3}$, which is well below the typical ratio of $10^{-1}$ \citep{Frank_2014}, even when we double this value to account for both jet lobes.

\begin{table}
  \caption{Line fluxes and ratios \label{tab:lines}}
  \begin{tabular}{cclc}
  \hline
  \hline
  \\[-0.2cm]
  \multicolumn{4}{c}{Line fluxes}\\[0.3cm]
  Line  & Nominal & Velocity & Flux\\
      & wavelength (\AA) & (km\,s$^{-1}$) & ($10^{-15}$\,erg\,s$^{-1}$\,cm$^{-2}$)\\
      \text{[O~{\sc iii}]} & 5007 & \multicolumn{1}{c}{--} & \multicolumn{1}{c}{$<0.02$} \\            
      \text{H~{$\beta$}}  & 4861 & $-171\pm6$ & $0.296\pm0.020$\\
      \text{[O~{\sc i}]} & 6300 & $-160\pm4$ & $0.090\pm0.006$ \\      
      \text{[N~{\sc ii}]} & 6548 & $-173\pm5$ & $0.075\pm0.005$\\
      \text{H~{$\alpha$}} & 6563 & $-171\pm1$ & $0.968\pm0.006$\\
      \text{[N~{\sc ii}]} & 6583 & $-168\pm2$ & $0.206\pm0.004$\\
      \text{[S~{\sc ii}]} & 6716 & $-155\pm18$ & $0.082\pm0.017$ \\
      \text{[S~{\sc ii}]} & 6731 & $-162\pm16$ & $0.069\pm0.017$ \\
      
  \hline\\[-.2cm]
  \multicolumn{4}{c}{Line ratios}\\[0.3cm]
  Numerator & Denominator & Ratio & log Ratio\\
  \text{[S~{\sc ii}]} 6716 & \text{[S~{\sc ii}]} 6731 & $1.19\pm0.38$ & 0.07\\
  \text{[N~{\sc ii}]} 6583 & \text{[O~{\sc i}]} 6300 & $2.29\pm0.16$ & 0.36\\
  \text{[S~{\sc ii}]} 6731 & \text{H}$\alpha$ 6563 & $0.07\pm0.02$ & -1.15 \\
  \text{H}$\alpha$ 6563 & {H}$\beta$ 4861 & $3.35\pm0.23$ & 0.53\\
  \text{[N~{\sc ii}]} 6583 & \text{[S~{\sc ii}]} 6716+6731 & $1.36\pm0.22$ & 0.13\\
  \text{[O \sc i]} 6300 & \text{H}$\alpha$ 6563 & $0.093 \pm 0.006$ & -1.03\\
  \hline
  \end{tabular}
\end{table}

\section{Discussion}
We clearly detect diffuse emission in typical jet lines at distances and blueshifts that are normal for protostellar jets from CTTSs and HAe/Be stars. Furthermore, the PA of the jet is perpendicular to the disk major axis \citep[PA$\approx137^\circ$;][]{Mendigutia_2017}, and a jet launched from the HD~100546 disk would be blueshifted at the PA of the diffuse emission given the disk inclination \citep[$i\approx44^\circ$ for the inner and outer disk, ][]{Mendigutia_2015}. We therefore interpret the diffuse emission as the protostellar jet from HD~100546. This jet emission went unnoticed in previous data of HD~100546 probably because the central star is very bright. This brightness requires short integration times to avoid saturation effects. However, jet features will remain hidden in the noise with short exposures. 
For our observations, we deliberately overexposed the star to increase the detectability of jet features. Another possibility to explain the nondetection of the jet in previous data is the sky coverage in spectroscopic data was not good enough.

The most prominent features of the MUSE data are the knot- and arc-like features. A collimated jet is not immediately recognizable in the MUSE data, which we attribute to the relative brightness of the bow shock and the jet body: the
jet emission is also far weaker than the bow shock in the prototypical HH\,1/2 system \citep{Hartigan_2011}. In addition, PSF spillover hinders the detection of jet emission close to the driving source. 
Still, the H$\alpha$
and [N~{\sc ii}]~$\lambda6583$ PVDs reveal that jet emission can be traced back to HD~100546 at essentially the same velocity as farther out, so that we are confident that HD~100546 is the driving source of the jet.

We note that the nondetection of the counter jet is not surprising: one jet lobe is typically brighter and sometimes also less collimated \citep{Hirth_1994}, so that its emission is likely lost in the region close to the driving source. In the specific case of the HD~100546 MUSE data, the counter jet should be visible if it were of comparable brightness as the forward jet, within the MUSE field of view (approximately~27 arcsec or $\sim3000$\,au in the jet direction, see Fig.~\ref{fig:full_overview}), and beyond the outer radius of the disk \citep[260\,au and 400\,au for the dust and gas disk, respectively; see][for a discussion of the HD~100546 disk sizes]{Pinila_2015}.

\subsection{Jet properties and relation to disk structure}

The jet axis aligns well with the bar-like feature (PA$\sim45^\circ$) seen in polarized H$\alpha$ and continuum light detected by \citet[][]{Mendigutia_2017}, 
which can be traced down to about 2\,au from the position of HD~100546; this is about the size of the inner disk.
\citet[][]{Mendigutia_2017} previously remarked on the similarity of this feature to a bipolar outflow. We are therefore tempted to associate the bar with the base of the jet. \citet{Mendigutia_2017} discarded the jet explanation because no jet was known for HD~100546 and no blueshifted excess emission was seen in [O~{\sc i}]~$\lambda6300$ \citep{Acke_2006}, which is a typically feature in jet-driving CTTSs.
The CO emission of HD~100546 also appears to require non-Keplerian disk emission in the bar region \citep{Walsh_2017}, which has been interpreted in terms of a warped disk and radial flows, but may instead be associated with the outflow detected here.

The association of the bar-like feature with the base of the jet is particularly interesting because it provides a strong constraint for the launching region ($\lesssim2\,$au). Such a small launching region  is generally consistent with 
other observational constraints such as jet rotation measurements and their upper limits, which locate the launching region within about 1\,au  for most CTTS jets \citep{Ferreira_2006, Coffey_2015}. This also agrees with the interpretation of high-resolution spectra
of CTTSs that suggest a jet origin within a few au from the star for all kinematic components \citep[low-velocity broad or narrow components and high-velocity components;][]{Fang_2018}.

A small launch radius is also compatible with jet models: they postulate that outflows that collimate into jets are launched within $\sim10$\,au from the driving source \citep{Pudritz_2007} such as magnetospheric ejections \citep{Zanni_2013, Romanova_2016} 
and some disk-wind models in which  the highest mass-loss occurs within about 1\,au \citep[for CTTSs, see][]{Nolan_2017}. On the other hand, the  significant mass-loss out to radii of 10--20\,au present in some magnetohydrodynamical simulations of full disks \citep[e.g.,][]{Bai_2014, Gressel_2015, Bethune_2017} 
is neither seen in our MUSE nor in the SPHERE data of HD~100546.
Outflows launched from larger radial distances are expected to be slower, and \citet{Fang_2018} suggested that they might be invisible in the classical 
jet lines because their temperature is low ($<5,000$\,K).
Specifically for HD~100546, however, slower outflow components would still be fast enough to shock heat material sufficiently so that it radiates in the typical jet lines. The lack of slow jet components in HD~100546 might therefore  also be rooted in the disk gap \citep[including gas depletion, see][]{Plas_2009} because this would also lower the mass-loss rate compared to the rate in full disks.
Without angular momentum removal by the jet, radial flows are suppressed and no or only little material would be transported through the gap, so that stellar accretion may be limited to the material in the 
inner disk. In this scenario, HD~100546 would be about to cease accretion within a few centuries \citep[][who reported a total inner disk mass of $5\times10^{-5}\,M_\odot$]{Mendigutia_2015, Pineda_2019}.

\subsection{Comparison to other jets}
The HD~100546 jet shows many features seen in other protostellar jets. First,  the emission is concentrated into rather discrete entities (knots). A (semi-) regular chain of emission knots (as seen, e.g., for HD~163296) is not detected in the MUSE data, however. This might be related to detection thresholds because additional emission knots might remain undetected at larger distances because the surface brightness might be too low, and closer to the star, the contrast against the PSF wings might be too low.

Second, the projected and deprojected velocities of $\approx170$\,km\,s$^{-1}$ and  230\,km\,s$^{-1}$ fall well within the velocity range of most atomic protostellar jets \citep[][the deprojected velocity assumes that the jet is launched perpendicular to the disk]{Eisloeffel_1998,Frank_2014}. The jets of HD~163296 and LkH$\alpha$\,233, driven by similarly massive central stars, show very similar velocities \citep{Ellerbroek2014, Melnikov_2008}.
Compared to  jets in Orion and Carina \citep{ODell_2003, Reiter_2017}, the HD~100546 jet is fast, but it is reasonable to assume that the apparent deceleration of the HD~100546 jet between about 18 and 21\,arcsec (Fig.~\ref{fig:PVDs}) would lead to similar velocities if it were observed at larger distances, as are most of the Orion and Carina jets.

Third, the density $n_e$ compares well with that of HD~163296, but is lower than that of other jets, such as the jet of LkH$\alpha$\,233. Still, densities derived for other jets are within about one order of magnitude  at similar distances because the highest densities are typically found close to the driving source. We do not have a significant detection or density diagnostics for the HD~100546 jet at short distances.

The mass-loss rate is on the very low end compared to the typical outflow rates of $10^{-7}$ to $10^{-9}\,M_\odot$\,yr$^{-1}$ measured for CTTSs \citep{Frank_2014}, and new observations of HD~100546 are required to determine whether this also holds for the jet close to the launch site. In particular, the measured outflow-to-accretion rate is lower ($3\times10^{-3}$ for the blueshifted lobe)
than typical ratios, which are usually  between  $10^{-2}$ and $10^{-1}$ \citep[see][]{Frank_2014}, also for the
brighter knots in the HD~163296 jet and likely for LkH$\alpha$\,233\footnote{Precise mass accretion rates are not available for LkH$\alpha$\,233, but the estimated outflow rate of $10^{-8}\,M_\odot$\,yr$^{-1}$ would result in $M_{out}/M_{acc}\gtrsim0.1$ for typical stellar accretion rates.}.
Taken together, the HD~100546 jet in the atomic lines traced by the MUSE data appears to be less powerful than other jets, both in absolute terms and in relation to the stellar accretion rate.

\section{Conclusions}
The presence of a collimated jet in a nonmagnetic or weakly magnetic 
relatively massive and evolved star with  a small inner disk provides immediate constraints on the jet-launching mechanism. First, the similarity in appearance between the HD~100546 jet and CTTS and other HAe/Be jets suggests a similar launching mechanism, that is, an origin within the innermost au from the central star. \citet{Mendigutia_2015} found that HD~100546 likely accretes magnetospherically (in contrast to boundary layer accretion), and the weak stellar magnetic field ($\lesssim100$\,G) may be sufficient to make HD~100546 a scaled-up version of a CTTS, including its jet. Second, regardless of what causes the formation of disk gaps or spirals, the jet-launching mechanism is left intact. Finally, planet formation and jets are not mutually exclusive if any of the planetary candidates in HD~100546 is real.

Herbig Ae/Be stars are by definition relatively evolved objects compared to many CTTSs, so that the accretion as well as the outflow rate are expected to have decreased compared to younger systems; specifically, HD~100546 has a quoted age of $4.8^{+2.0}_{-0.2}$\,Myr  \citep{Wichi_2020, Acke_2006}, which is older than the typical CTTS age of 1--2\,Myr. 
We may speculate that the low outflow rates at continued high accretion rates imply that mass-outflow rates decline before the accretion drops. 
\citet{Fang_2018} found that the occurrence frequency of the high-velocity component in typical jet lines is significantly lower 
in systems with transitional disks than in full disks, while stellar accretion rates may be very similar \citep{Manara_2014}. Within this picture, the low mass-loss rate of the HD~100546 jet may indicate an evolutionary sequence with jet mass-loss rates that decrease first and stellar accretion rates that decrease only later, in  the phase when the residual material in the 
inner disk is consumed and the supply of disk material from outer disk radii is reduced or absent. If the disappearance of the jet causally changes the disk, for example, by reducing radial transport through the disk, or if jet activity ceases, for example, as a result of other processes such as gap formation needs to be investigated in future work.

\begin{acknowledgements}
PCS gratefully acknowledges support by DLR 50 OR 1901. HMG acknowledges support by NASA-HST-GO-12315.01. 
NH has been partially funded by the Spanish State Research Agency (AEI) 
Project No. ESP2017-87676-C5-1-R and No. MDM-2017-0737 Unidad de 
Excelencia 'Mar\'{\i}a de Maeztu'- Centro de Astrobiolog\'{\i}a (INTA-CSIC).
Support for this work was provided by the National Aeronautics and
Space Administration through Chandra Award Number GO5-16014X issued by
the Chandra X-ray Observatory Center, which is operated by the
Smithsonian Astrophysical Observatory for and on behalf of the National
Aeronautics Space Administration under contract NAS8-03060.
\end{acknowledgements}

{}

\begin{appendix}

\section{Data reduction details \label{sect:datareduction}}
Data reduction generally followed standard procedures. However, the immediate region around the central star is heavily affected by saturation effects. The particular sub-IFUs of the 24 individual sub-IFUs of MUSE, which are most strongly affected by the  central star, were discarded during further data reduction. Each of the 24 sub-IFUs covers a particular patch on the sky, and discarding one to two sub-IFUs creates empty sky patches for each exposure. Full sky coverage\footnote{Except for about 2 arcsec around the central star} is then obtained by rotating the detector by 22.5~deg between individual exposures, as is generally recommended to improve spatial homogeneity, and then combing the individual exposures.

\begin{figure}
  \centering
  \includegraphics[width=0.49\textwidth]{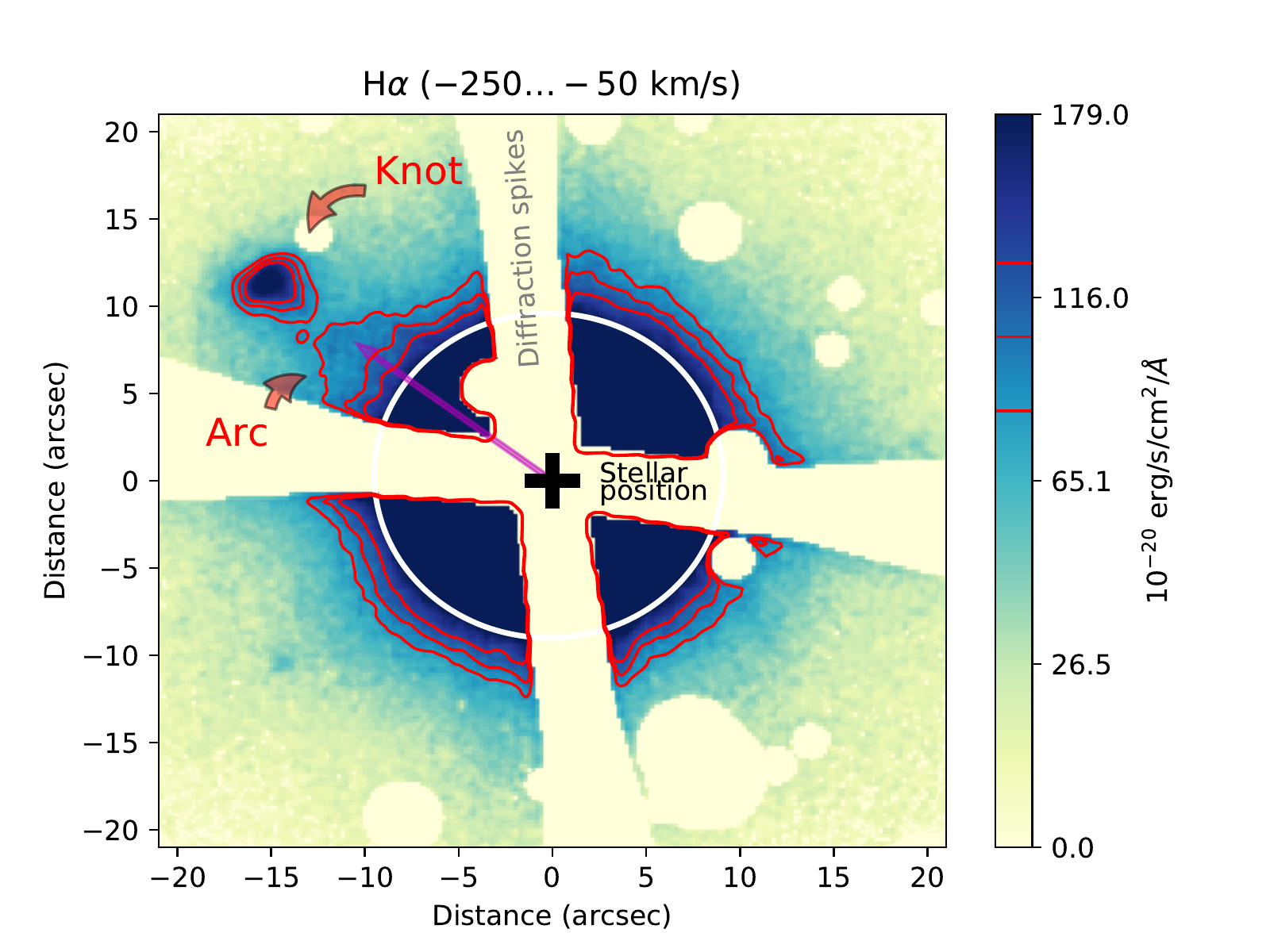}
  \caption{Most of the field of view around HD~100546 in H$\alpha$ before the stellar emission was removed. The knot and arc are already visible before PSF subtraction. The white circle shows the symmetry of the stellar PSF.}
  \label{fig:full_overview}
\end{figure}

Point-spread function subtraction is performed by considering the PSF as a combination of three radially symmetric exponential functions. Moffat functions, typically used to describe the geometrical form of a PSF, were found to perform less well in the outer parts of the PSF ($d\gtrsim 5$ arcsec). 
PSF fitting (or rather the PSF wing fitting) was performed individually for several wavelength regions containing one to several relevant jet emission lines. During the fitting procedure, background stars, the diffraction spikes, and the sky patches containing jet emission were discarded. For each relevant emission line, the PSF fit was then performed to a cube restricted to wavelengths around the line centroid (approximately 30\,\AA,{}  excluding the -400 to +100 \,km\,s$^{-1}$ range 
around the nominal wavelength). The fit assumes that the PSF has the same wavelength dependence as the stellar spectrum extracted from an unsaturated region close to the stellar position to minimize the number of free parameters. The PSF model ensures spatial and spectral smoothness, requires only a small number of free parameters, and was found to leave systematic PSF subtraction residuals only very close to the star and the diffraction spikes (see Fig.~\ref{fig:full_overview}).

\section{Alternative mass-loss estimate \label{sect:alt_mass_loss}}
When we used Eq.~A8 in \citet{Hartigan_1995} or the numerical constants used by \citet{Fang_2018}, we obtained similar values (differences $\lesssim15\,$\%) as those described in the main text. Also, the alternative method to estimate the mass-loss rate based on area $A$, velocity
$v$, and density estimates ($n$) results in comparable values:
\begin{equation}
\dot M = \rho v A = \mu m_H n v A \ , \label{eq:Mdot_alt}
\end{equation}
where $\rho= \mu m_H n$ is the mass density with the mean
molecular weight $\mu = 1.24$, the mass of the hydrogen atom $m_H$, and
the number density $n$. To estimate $n$ from $n_e$ derived from the [S~{\sc ii}] doublet, \citet{Hartigan_1994} recommend to use 
\begin{equation}
n = \frac{n_e}{\sqrt{<C>}<I>}\,, \label{eq:n}
\end{equation}
with the ionization $<I>$ fraction and the mean compression factor $<C>$.

The HD~100546 jet appears to be of low density ($n_e\sim250$\,cm$^{-3}$) and to have a high ($\sim80$\,km\,s$^{-1}$) shock velocity. Therefore, 
$\log <C> \sim 2$ (cf. Fig.~17 in \citet{Hartigan_1994}), while $<I>$ may be around 0.5. Equation~\ref{eq:n} therefore indicates that $n\sim50$\,cm$^{-3}$ (for $n_e=250$\,cm$^{-3}$), and Eq.~\ref{eq:Mdot_alt} gives $\dot{M} \sim 7\times10^{-10}$\,$M_\odot\,$yr$^{-1}$. This is reasonably close to the value provided in Sect.~\ref{sect:results}.
%
%

\end{appendix}


\end{document}